\documentclass[a4paper]{jpconf}

\usepackage{graphicx}
\usepackage{amssymb}
\usepackage{amsthm}
\usepackage{amsmath}
\usepackage{mathrsfs}  
\usepackage{bm}
\usepackage{epsfig}
\usepackage{stackrel}

\newcommand{\mcN}{{\mathcal N}}

\newcommand{\ren}{{\mathrm{ren}}}
\newcommand{\reg}{{\mathrm{reg}}}
\newcommand{\ct}{{\mathrm{ct}}}
\newcommand{\GH}{{\mathrm{GH}}}
\newcommand{\QCD}{{\mathrm{QCD}}}

\begin{document}
\title{Thermodynamics of AdS$_5$ black holes: holographic QCD and St\"uckelberg model}

\author{Eugenio Meg\'{\i}as$^{1,2}$}

\address{$^{1}$Departamento de F\'{\i}sica At\'omica, 
Molecular y Nuclear and Instituto Carlos I de F\'{\i}sica Te\'orica y Computacional, Universidad de Granada, Avenida de Fuente Nueva s/n, E-18071 Granada, Spain
}
\address{$^{2}$Departamento de F\'{\i}sica Te\'orica, 
Universidad del Pa\'{\i}s Vasco UPV/EHU, Apartado 644, E-48080 Bilbao, Spain}

\ead{emegias@ugr.es}

\begin{abstract}
We explore the thermodynamics of AdS$_5$ black holes in two models: i)
an improved holographic QCD model with a simple dilaton potential, and
ii) the St\"uckelberg model in 5D. In the former case, by applying
techniques of singular perturbation theory, we obtain a resummation of
the naive expansion at high temperatures, providing a good fit to the
lattice data for the trace anomaly. In the latter, we find a solution
of the equations of motion by considering an expansion in the
conformal dimension of the current associated to the gauge~field.
\end{abstract}

\section{Introduction}
\label{intro}

The gauge/gravity duality is a powerful tool to study the properties
of gauge theories, and in particular of QCD, in their strongly coupled
regime. Using this duality, the thermodynamics of a field theory can
be obtained from the classical computation of the thermodynamics of
black holes in the gravity dual. The entropy of a black hole can be
computed from the famous Bekenstein-Hawking entropy
formula~\cite{Gubser:2008ny,Gursoy:2008za,Alanen:2009xs,Megias:2010ku,Li:2011hp}. In
conformal AdS$_5$ the entropy scales like $S_{\textrm{Black Hole}}
\propto r_h^3 \propto T^3$; however, in order to have a reliable
extension of this duality to SU($\text{N}_c$) Yang-Mills theory, the
first task is to control the breaking of conformal
invariance. Conformal transformations are broken by quantum
corrections due to the necessary regularization of the UV divergences,
yielding the so-called {\it trace
  anomaly}~\cite{Collins:1976yq,Landsman:1986uw}
\begin{equation}
\langle T^\mu_\mu\rangle = -\varepsilon + 3p = -\frac{\beta(g)}{2g^3} \langle (F_{\mu\nu}^a)^2\rangle  \,,
\end{equation}
where $T^{\mu\nu}$ is the energy-momentum tensor, $\varepsilon$ is the
energy density, $p$ is the pressure and $\beta$ is the beta
function. We will explore the application of techniques
of singular perturbation theory to get a better control of the high
temperature expansion of the thermodynamic quantities.

On the other hand, theories with a massive gauge field in the bulk of AdS$_5$ have been proposed as holographic duals of QCD with dynamical gauge fields. The relation of the St\"uckelberg field in holography to the axial anomaly has been first pointed out in Ref.~\cite{Klebanov:2002gr}. Recent applications of the St\"uckelberg mechanism include the computation of anomalous effective actions~\cite{Coriano:2007fw}, anomalous transport and their renormalization properties~\cite{Gursoy:2014ela,Jimenez-Alba:2014iia}, AdS/QCD~\cite{Casero:2007ae} and beyond the Standard Model scenarios~\cite{Kiritsis:2003mc}. In the last part of this work we will study the thermodynamics of the St\"uckelberg model in 5D as well as the consistency of the entropy formula in this model.

%---------------------------------------------------------------------------------
%---------------------------------------------------------------------------------
\section{The holographic QCD model}
\label{sec:model}

The bottom-up approach is based on the building of a gravity dual of
QCD, including the main properties of this theory. We introduce in this
section the model and study the solutions.

\subsection{The model}
\label{subsec:model}

One of the most successful holographic models of QCD within the
bottom-up scenario is the 5D Einstein-dilaton model, with the
Euclidean action~\cite{Gursoy:2008za}
\begin{equation}
\begin{split}
\label{eq:Smodel}
S &= \frac{1}{2\kappa^2} \int d\rho \, d^4x \sqrt{G} \Bigl(-R + G^{\mu\nu} \partial_\mu \Phi \partial_\nu \Phi + 2V(\Phi) \Bigr) + S_{\GH} \,,
\end{split} 
\end{equation}
where $\kappa^2 = 1/(2 M^3)$ is the 5D Newton constant with $M$ the
Planck mass, and $\Phi$~is a scalar field to be identified with the
Yang-Mills coupling through $g^2 = e^{\gamma \Phi}$. The boundary
term~$S_{\GH}$ is the usual Gibbons-Hawking contribution built up from
the extrinsic curvature. The introduction of a scalar field breaks
conformal invariance, and the form of the scalar potential $V(\Phi)$
is usually phenomenologically adjusted to describe some observables of
QCD. We can introduce the superpotential, whose relation with the
scalar potential is $V(\Phi) = \frac{1}{2} W'(\Phi)^2 - \frac{2}{3}
W(\Phi)^2$.  In this work we will consider the simple form of the
superpotential
\begin{equation}
W(\Phi) = \ell^{-1} (3 + v_0 e^{\gamma \Phi}) \,,
\end{equation}
where $\ell$ is the radius of the asymptotically AdS$_5$
background. The combination $\gamma^2 v_0/2 \equiv b_0$ can be taken
from the one-loop $\beta$-function of a pure SU($\text{N}_c$) gauge theory, i.e.
\begin{equation}
\beta(g) := \mu \frac{dg}{d\mu} = -b_0 g^3 =  -\frac{11 \, \text{N}_c}{3 (4 \pi)^2} g^3  \,. \label{eq:betag}
\end{equation}

\subsection{Equations of motion}
\label{subsec:eom}

We are interested in the study of finite temperature solutions of the
Einstein-scalar gravity model corresponding to a black hole. In Fefferman-Graham coordinates the metric is
 \begin{equation}
 ds^2 = G_{\mu\nu} dx^\mu dx^\nu = \frac{\ell^2}{4\rho^2} d\rho^2 + \frac{\ell^2}{\rho} g_{\tau\tau}(\rho) d\tau^2 +  \frac{\ell^2}{\rho} g_{xx}(\rho) d{\vec{x}}^2 \,.  \label{eq:metric} 
 \end{equation}
The asymptotic expansions for the metric and the dilaton field in
holographic QCD models have been studied in details in these
coordinates~\cite{Papadimitriou:2011qb}. This metric has a regular
horizon at $\rho=\rho_h$, i.e. $g_{\tau\tau}(\rho_h)=
g_{\tau\tau}'(\rho_h)=0$, and the temperature and entropy density of
the black hole are given~by
\begin{equation}
T = \frac{1}{2\pi} \sqrt{2\rho_h g_{\tau\tau}^{\prime\prime}(\rho_h)} \,, \qquad  s = \frac{2\pi}{\kappa^2} \left( \frac{\ell^2}{\rho_h } g_{xx}(\rho_h) \right)^{3/2} \,. \label{eq:TS}
\end{equation}
Using this metric, the equation for the scalar field, which is a
second order differential equation, can be decoupled by an additional
differentiation. Then it can be converted in autonomous by making the
change of variables, $(\rho, \Phi) \to (\Phi, u(\Phi))$, where
$u(\Phi) \equiv \rho \, \Phi'(\rho)$, leading to~\footnote{We refer the
  reader to Ref.~\cite{Megias:2017czr} for the remaining field
  equations involving $g_{xx}(\Phi)$ and $g_{\tau\tau}(\Phi)$.}
\begin{equation}
u''(\Phi) = \frac{\bigl(u'(\Phi) \bigr)^2}{u(\Phi)} - \frac{3 \ell^2 V'(\Phi)}{4 u(\Phi)^2} u'(\Phi) + 
\frac{\ell^4 \bigl(V'(\Phi) \bigr)^2}{16 u(\Phi)^3} + \frac{\ell^2 \bigl(8 V(\Phi) + 3 V''(\Phi) \bigr)}{12 u(\Phi)} \,. \label{eq:u}
\end{equation}
At zero temperature, when $g_{xx} = g_{\tau \tau}$, the solution corresponds  to a domain wall configuration,  $u_0(\Phi) = \frac{\ell}{2} W'(\Phi) = \tfrac{\gamma v_0}{2} e^{\gamma \Phi}$, and it is given by 
\begin{equation}
e^{\gamma \Phi(\rho)} = \frac{2}{ \gamma^2 v_0\,  \log \left(\rho_0/\rho \right) } \,,  \qquad g_{xx}(\Phi) = g_{\tau \tau}(\Phi) = g_{(0)} \cdot e^{-2 \Phi/(3 \gamma)}  \,.  \label{eq:zeroim}
\end{equation}
Using $g^2 = e^{\gamma \Phi}$ and Eq.~(\ref{eq:betag}), it turns out
that the radial coordinate $\rho^{-1/2}$ can be interpreted as the
renormalization group (RG) scale, and $\rho_0^{-1/2} \equiv
\Lambda_{\QCD}$ is an integration constant corresponding to the
location of the Landau pole of QCD.  The other integration constant
$g_{(0)}$ is arbitrary.

\subsection{Asymptotically AdS black hole solutions from resummation}
\label{subsec:Resummation}

We now consider a black hole solution specified by the horizon data
$(\rho_h, \Phi_h)$. Local analysis of the equations of motion shows
that, in order to have a regular horizon at $\Phi= \Phi_h$, it is
necessary that $\lim_{\Phi \to \Phi_h} \bigl(u(\Phi) g'_{xx}(\Phi) -
g_{xx}(\Phi)\bigr) = 0$, which fixes the first term of the power
series solution
\begin{equation}
u(\Phi) = \sum_{n=0}^\infty a_n(\Phi_h) (\Phi_h - \Phi)^{n + 1/2} \,, \label{eq:seriesu}
\end{equation}
and similar expressions for $g_{xx}(\Phi)$ and
$g_{\tau\tau}(\Phi)$. We will study the analytical solution in the
regime $\nu \equiv \gamma v_0 e^{\gamma \Phi_h} \ll 1$ for any $\gamma
>0$ by using a boundary-layer analysis. In this regime the leading
part of each coefficient of the series (\ref{eq:seriesu}) behaves as
$a_n(\Phi_h) \propto \nu^{-n + 1/2}$. This suggests to consider an
expansion of the form $u^{\mathrm{in}}(\Phi) = \sum_{n=1}^\infty \nu^n
U_n((\Phi_h - \Phi)/\nu)$ which is valid in the region of
boundary-layer near the horizon, where the inner variable, $\Psi
\equiv (\Phi_h - \Phi)/\nu$, is $O(1)$. The first order of this
expansion reads $u^{\mathrm{in}}(\Phi) = \frac{\nu}{2} \sqrt{1 - e^{-
    4 \Psi}} + O(\nu^2)$. For the regime outside the boundary layer we
tried a solution of the form $u^\mathrm{out}(\Phi) = \sum_{n=1}^\infty
\nu^n u_n(\Phi_h - \Phi)$, and the corresponding substitution in
Eq.~(\ref{eq:u}) leads to $u^\mathrm{out}(\Phi) = \frac{\nu}{2}
e^{\gamma (\Phi - \Phi_h)}$ to all orders in $\nu$, so that the outer
solution has the same form as the zero temperature solution. Thus the
asymptotic matching produces a uniform approximation valid for
$(-\infty, \Phi_h)$ given by
\begin{equation}
u^{\mathrm{unif}}(\Phi) = \frac{v_0 \gamma}{2} e^{\gamma \Phi} + u^{\mathrm{in}}(\Phi) -\frac{v_0 \gamma}{2} e^{\gamma \Phi_h} + O(\nu^2) \,, \label{eq:unif}
\end{equation} 
so that $|u(\Phi) - u^{\mathrm{unif}}(\Phi)| = O(\nu^2)$. The last
term in Eq.~(\ref{eq:unif}) is the common limit of the inner and outer
approximation in the matching region, $\nu \ll \Phi_h - \Phi \ll 1$.

From the knowledge of the lowest orders of $u^{\mathrm{in}}(\Phi)$, it
is possible to get in closed form the lowest orders in $\nu$ for the
inner solutions of the metric, $g_{i j}^{\mathrm{in}}(\Phi) = \sum_n
\nu^n g_{(n) i j}(\Psi)$, while the outer solution reduces to the
zero temperature result given by Eq.~(\ref{eq:zeroim}).  Finally, the
asymptotic matching of the outer solution with the inner expansion
determines the horizon quantities
\begin{align}
g_{x x}(\Phi_h) &=  2 g_{(0)} e^{-2 \Phi_h/(3  \gamma)} \left(1 -\frac{\nu \, \log 2}{3 \gamma} +  O(\nu^2)  \right) \,, \label{eq:ders} \\
g_{\tau \tau}^\prime(\Phi_h) &= 
 -\frac{8 g_{(0)} e^{-2 \Phi_h/(3 \gamma)}}{\nu}  \left( 1 + \frac{\nu}{12 \gamma} (4 + 3\gamma^2 - 4\log 2)   + O(\nu^2)  \right) \,. \label{eq:der1}
\end{align}
Note that the second derivative in Eq.~(\ref{eq:TS}) is written as
$g_{\tau \tau}''(\rho_h) = \rho_h^{-2} \lim_{\Phi \to \Phi_h}
\bigl(u(\Phi) u'(\Phi) g_{\tau \tau}'(\Phi) \bigr)$, so that these
quantities are directly related to the temperature and entropy of the
black hole.

%%%%%%%%%%%%%%%%%%%%%%%%%%%%%%%%%%%%%%%%%%%%%%%%%%%%%%%%
\section{Thermodynamics of holographic QCD}
\label{sec:thermodynamics}
%%%%%%%%%%%%%%%%%%%%%%%%%%%%%%%%%%%%%%%%%%%%%%%%%%%%%%% 

Using the previous results, we can apply the holographic prescription~\cite{Klebanov:1999tb} for the derivation of the trace anomaly of the Yang-Mills theory (see e.g. Ref.~\cite{Gursoy:2008za} for full details on the procedure).

\subsection{Trace Anomaly}
\label{subsec:TraceAnomaly}

The perturbative running of the coupling constant in Yang-Mills theory induces a deformation in the action which may be written as 
\begin{equation}
\delta S = \delta\left( \frac{1}{2 g^2} \right) \int  d{^4} x  \, \Tr F^2 = \int d{^4} x \left(\frac{\gamma}{2} e^{-\gamma \Phi} \delta \Phi \right) \left(-\Tr F^2 \right) \,.  \label{eq:break}
\end{equation}
This gives rise to the trace anomaly of the stress tensor 
\begin{equation}
T^\mu_\mu = - \frac{\beta(g)}{g^3} \Tr F^2 = b_0 \Tr F^2 \,.
\end{equation}
To apply the holographic prescription we can assume that the boundary
value of the combination $\frac{\gamma}{2 b_0} e^{-\gamma \Phi(\rho)}
\delta \Phi(\rho)$ plays the role of a source that couples to the dual
operator $\mathcal{O} = -b_0 \Tr F^2$. The one-point function $\langle
\mathcal{O} \rangle$ is therefore proportional to the gluon
condensate.

With the zero temperature solution $\Phi_0(\rho)$ given by
Eq.~(\ref{eq:zeroim}) and $\gamma^2 v_0/2=b_0$, one can prove that a
change in the scale $\delta \rho_0$, induces a change in the scalar
field $\delta \Phi$ of the form~$\frac{\gamma}{2 b_0} e^{-\gamma
  \Phi_0} \delta \Phi = - \frac{\delta \rho_0}{2 \rho_0}$. Then,
according with Eq.~(\ref{eq:break}), the derivative of the free energy
density $w$ with respect to $\log \rho_0$ must be proportional to the
gluon condensate, leading to
\begin{equation}
2 \rho_0 \frac{\partial w}{\partial \rho_0} = b_0  \langle \Tr F^2 \rangle  \qquad \textrm{where} \qquad  S = \frac{V_3}{T} \, w \,. 
\end{equation}

\subsection{First law of thermodynamics}
\label{subsec:1stLaw}

We then assume that the free energy density is a function of the source
$\rho_0$ and the temperature~$T$, whose variation is $d w = -s\, dT -
\langle \mathcal{O} \rangle d\rho_0/(2 \rho_0)$. This assumption
requires the condition for integrability $2 \rho_0 \partial s /
\partial \rho_0 = \partial \langle \mathcal{O} \rangle / \partial T$,
which has been checked in Ref.~\cite{Megias:2017czr} by using
holographic renormalization techniques. Since there is no conserved
charge, the Euler identity adopts the form $w = \varepsilon -T
s$. Then, it follows that the first law of thermodynamics is
\begin{equation}
d\varepsilon = T \, ds - \langle \mathcal{O} \rangle \frac{d\rho_0}{2 \rho_0} \,. 
\end{equation} 
At this point one can observe that there is a similarity to the
thermodynamic identities of elastic bodies, where the work done by an
applied stress $\tau_{i j}$ is given by $dW = -V_3 \tau_{i j}\, d\eta_{i
  j}$, being $\eta_{i j}$ the strain deformation and $V_3$ the volume,
see e.g. Refs.~\cite{landau,wallace}.  This means that the source
$\sim \log \sqrt{\rho_0}$ plays the role of the deformation, while the
condensate $ \langle\mathcal{O} \rangle$ is the analogous of the
stress tensor. There is no need to interpret the source as any
kind of chemical potential~\cite{Buchel:2003ah,Lu:2014maa}, and this
analogy with the thermodynamics of deformation seems more suitable.

\subsection{Equation of state when $\nu \ll 1$}
\label{subsec:EoS}

\begin{figure}[t]
\begin{center}
\epsfig{figure=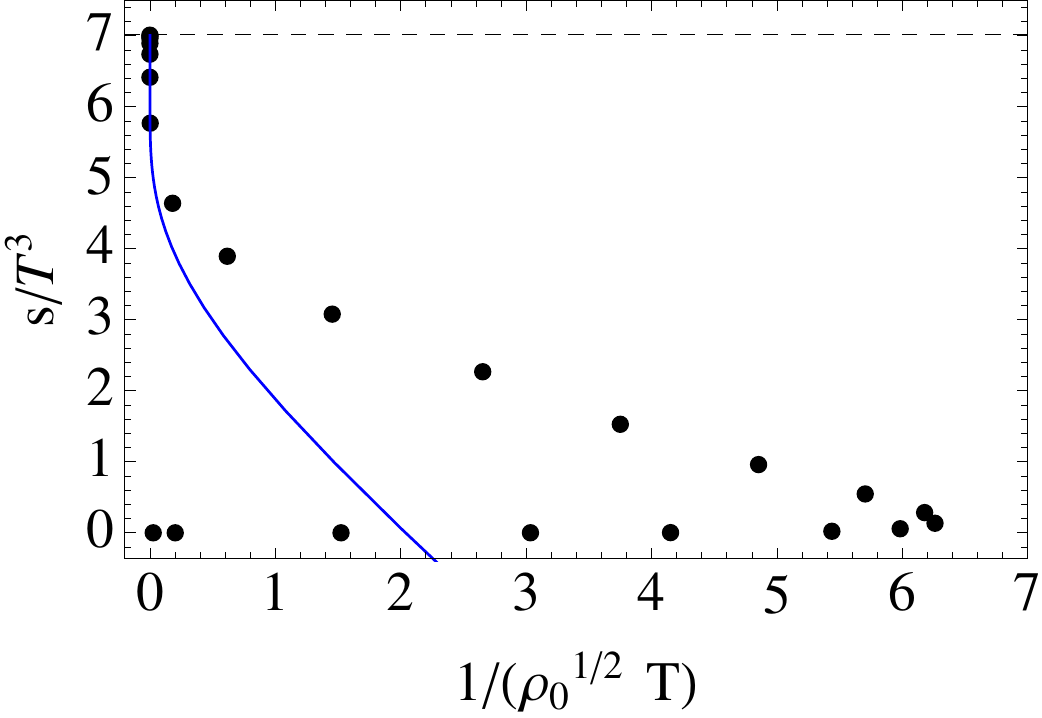,width=7.1cm} \hspace{1.5cm} \epsfig{figure=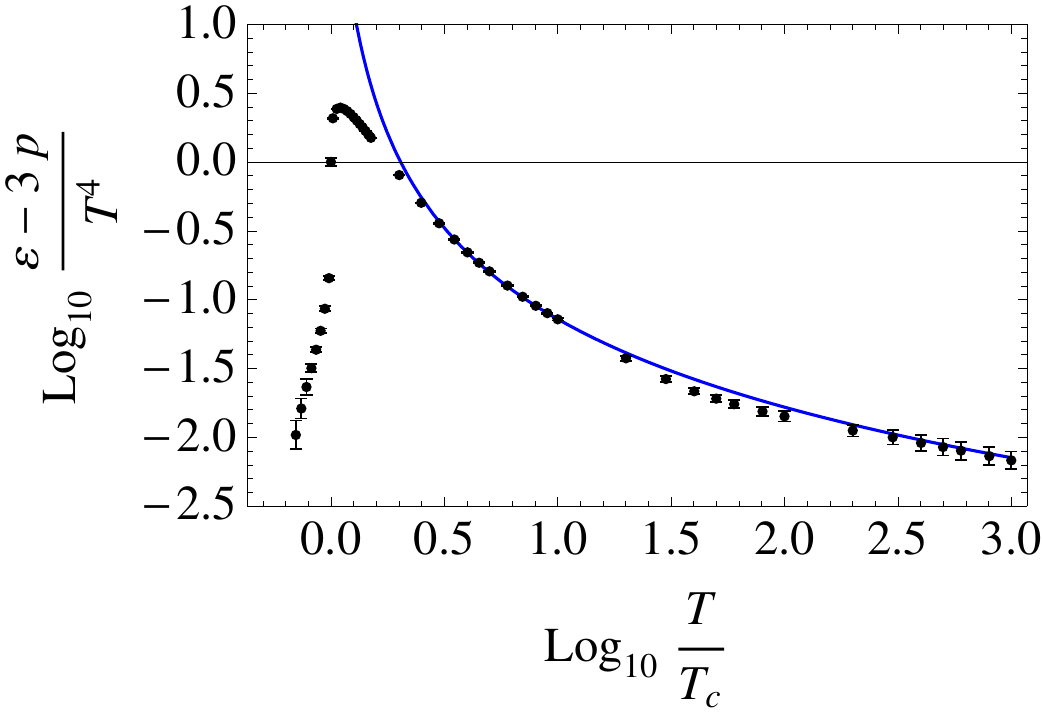,width=7.1cm}
\end{center}
\vspace{-0.5cm}
\caption{Left panel: $s/T^3$ as a function of~$(\rho_0^{1/2}
  T)^{-1}$. We display as dots the result from a numerical computation
  of the equations of motion, while the solid (blue) line corresponds
  to the analytical result of Eq.~(\ref{eq:sT3}). The horizontal
  dashed lines correspond to the Stefan-Boltzmann limit. We have used
  $\text{N}_c=3$ and $\gamma = \sqrt{2/3}$. Right panel:
  $(\varepsilon-3p)/T^4$ as a function of $T$.  The dots correspond to
  the lattice data for gluodynamics from \cite{Borsanyi:2012ve} with
  $\text{N}_c=3$. We show as a solid (blue) line the analytical result
  from Eq.~(\ref{eq:e3p}) with the values of the parameters given in
  Sec.~\ref{subsec:lattice}. }
\label{fig:entropy_traceanomaly}
\end{figure}

Using dimensional analysis one finds that a compact way of writing the
pressure is
\begin{equation}
p(T, \rho_0)  = - w = a T^4 + a T^4 \, \mcN(T^2 \rho_0) \,, 
\end{equation} 
where $a$ is a constant and $\mcN(x)$ is a function to be determined. We
may now compute the pressure from the information about the horizon
encoded in Eqs.~(\ref{eq:ders}) and (\ref{eq:der1}), which determine
the temperature and entropy density when $T \sqrt{\rho_h} \gg 1$ and
$\Phi_h \to -\infty$. When using the relation $\rho_h \simeq 2 \rho_0
\exp\left( - \frac{2}{\gamma^2 v_0} e^{-\gamma \Phi_h} \right)$, one
obtains for the entropy density~\footnote{If one chooses the $5$D
  Newton constant $\kappa^2$ to reproduce the Stefan-Boltzmann limit
  of the entropy density in gluodynamics at high temperatures,
  $s_{\textrm{gluons}} \to (\text{N}_c^2-1)\frac{4\pi^2}{45}T^3$, then
  one has $\frac{\ell^3}{\kappa^2} = (\text{N}_c^2-1)
  \frac{2}{45\pi^2}$.}
\begin{equation}
\frac{s}{T^3} =  T^{-3} \frac{\partial p}{\partial T} = 4 a + 4 a  \, \mcN(T^2 \rho_0) + 2 a\, T^2 \rho_0 \, \mcN^\prime(T^2 \rho_0) \,, \label{eq:sT3}
\end{equation}
with $a = \ell^3 \pi^4/(2 \kappa^2)$ and $\mcN(x) =
-\frac{3}{\mathcal{W}\left( 3 v_0^{-1} (g_{(0)}^{-1} \pi^2 x)^{3
    \gamma^2/2} \right)} +O(\mathcal{W}^{-2})$, where $\mathcal{W}(z)$
denotes the principal branch of the Lambert
$\mathcal{W}$-function. The trace of the thermal stress tensor can be
obtained from the relation, $\varepsilon + p = T s = T \partial_T p$,
as
\begin{equation}
\varepsilon - 3 p = -b_0  \langle \Tr F^2 \rangle =  \frac{\ell^3 \pi^4 \gamma^2 T^4}{2 \kappa^2}  \mcN(T^2 \rho_0)^2 + \ldots \,. \label{eq:e3p}
\end{equation} 
The asymptotic behavior $\mathcal{W}(z) \sim \log z$ as $z \to \infty$
is consistent with the logarithmic behavior of the trace anomaly at
high temperature. Note that in the opposite limit, $\mathcal{W}(z)
\approx z$ as $z \to 0$, the model seems to predict the existence of
power corrections in temperature. Although these corrections are
desirable, as we will see below they don't play any phenomenological
role within this model. However, it is remarkable that the resummation
performed above leads to such a result. Finally, we display in the
left panel of Fig.~\ref{fig:entropy_traceanomaly} the entropy density
(normalized to $T^3$) as a function of the inverse of
temperature. Note that this is a multivalued function with two regimes
corresponding to big (upper branch) and small (lower branch) black
holes.

%%%%%%%%%%%%%%%%%%%%%%%%%%%%%%%%%%%%%%%%%%%%%%%%%%%%%%%%
\subsection{Phenomenological consequences for QCD}
\label{subsec:lattice}
%%%%%%%%%%%%%%%%%%%%%%%%%%%%%%%%%%%%%%%%%%%%%%%%%%%%%%% 

We can now compare the results of the model by considering $\gamma$
and $\rho_0$ as free parameters, with the lattice data of the equation
of state of gluodynamics in SU(3). The best fit to the lattice data of
the trace anomaly of Ref.~\cite{Borsanyi:2012ve} for temperatures $3
T_c \le T \le 1000 T_c$ leads to $\gamma = 2.375(19)$ and
$\rho_0T_c^2/g_{(0)} = 0.0555(10)$, with $\chi^2/\textrm{dof} =
0.088$. The result for the trace anomaly is displayed in the right
panel of Fig.~\ref{fig:entropy_traceanomaly}. We conclude that the
model leads to an accurate description of the lattice data in the
whole regime $3 \leq T/T_c$. For temperatures closer to $T_c$, it has
been discussed in the literature that a good description of the
lattice data requires the existence of power corrections in $T^2$, see
e.g. Refs.~\cite{Pisarski:2006yk,Megias:2009mp}. However, from this
analysis it seems to be that the model does not lead to power
corrections in this regime. In either case, we cannot exclude that
power corrections could become relevant after considering the effects
of RG flows connecting two fixed points, a study already outlined in
Ref.~\cite{Megias:2017czr}, or in a sophisticated version of the
model.

\section{The St\"uckelberg model}
\label{sec:Stueckelberg}

In the last part of this work we will study the thermodynamics of the St\"uckelberg model in 5D. This model is an extension of the Einstein-Maxwell theory including a massive gauge field.

\subsection{The model}
\label{subsec:Stueckelberg_model}

The St\"uckelberg model in 5D is defined by the Euclidean action
\begin{eqnarray}
S = \int dz d^4x \sqrt{G} \bigg[ \frac{1}{2\kappa^2} \left( R + \frac{12}{\ell^2} \right)  - \frac{1}{4 g_5^2} F_{\mu\nu} F^{\mu\nu} - \frac{1}{2 g_5^2} G^{\mu\nu} (\partial_\mu \Phi - m A_\mu)  (\partial_\nu \Phi - m A_\nu) \bigg] + S_{\GH} \,,
\end{eqnarray} 
where $A_\mu$ is a U($1$) gauge field with mass $m$. The mass of the gauge field is related to the conformal dimension $\tilde\Delta$ of the U($1$) current operator $J^\mu$ dual to the gauge potential as~$m^2\ell^2 = (\Delta - 3)(\Delta-1)$ with $3 \le\Delta < 4$, where for convenience we have defined $\Delta := 3 - \tilde\Delta$. Let us consider for the moment the metric in conformal coordinates
\begin{equation}
ds^2 = \frac{\ell^2}{z^2} \left[ dz^2 + g_{ij} dx^i dx^j\right] \,,
\end{equation}
which is asymptotically AdS$_5$ near the boundary, $z\to 0$. The behavior of the time component of the gauge field in this regime is
\begin{equation}
A_\tau(z) = A_{(0)\tau} \left( \frac{z}{\ell} \right)^{3-\Delta} + {\cal O}^\tau_A \left( \frac{z}{\ell} \right)^{\Delta-1} + \cdots \,, \label{eq:Anb}
\end{equation}
where $A_{(0)\tau}$ corresponds to a deformation of the conformal field theory (CFT), with energy scale $\Lambda = 1/\ell$, i.e.
\begin{equation}
\int d^4x \, A_{(0)\tau} {\mathcal J} = \int d^4x \, \Lambda^{3-\Delta} A_{(0)\tau} {\cal O}^\tau_A \,.
\end{equation}
In this expression ${\cal O}^\tau_A \equiv {\mathcal J} \Lambda^{\Delta-3}$ is a condensate of dimension~$\Delta$, while $A_{(0)\tau}$ plays the role of a chemical potential.

\subsection{Black hole solution}
\label{subsec:black_hole}

The equations of motion of the model follow from the variation of the action with respect to scalar and gauge fields, and the metric. Since we are interested in black hole solutions, we will consider the following ansatz
\begin{equation}
ds^2 = \frac{r^2 f(r)}{\ell^2} d\tau^2 + \frac{\ell^2}{r^2 g(r)} dr^2 + \frac{r^2}{\ell^2} d{\vec{x}}^2  \,, \qquad A(r) = j(r) d\tau \,, \qquad \Phi(r) \; = \; \textrm{const} \,,  \label{eq:dsBH} 
\end{equation}
where we have used $r = \ell^2/z$, and seek for perturbative solutions to first order in $\Delta-3$. The leading order solution, $\Delta=3$, corresponds to the well-known result in the Einstein-Maxwell theory in 5D, which reads
\begin{equation}
f_0(r) = g_0(r) \; = \; \left( 1 - \frac{r_h^2}{r^2}\right) \left( 1 + \frac{r_h^2}{r^2} - \frac{q^2}{r_h^2 r^4} \right) \,,  \qquad j_0(r) = A_{(0)\tau} \left( 1 - \frac{r_h^2}{r^2} \right) \,,
\end{equation}
where the charge associated to the U($1$) symmetry is
$q=\sqrt{\frac{2}{3}} \frac{A_{(0)\tau} \ell r_h^2 \kappa}{g_5}$, and
the temperature of the black hole is $T_0 = \frac{r_h}{\pi \ell^2}
\left( 1 - \frac{q^2}{2 r_h^6} \right)$. We use the parametrization
\begin{equation}
f(r) = f_0(r) \left(1 + F(r) \right) \,, \qquad g(r) = f_0(r) \left(1 + F(r) + G(r) \right) \,, \qquad j(r) = j_0(r) + J(r) \,,
\end{equation}
where $F,\; G, \; J$ are corrections of $O(\Delta-3)$. By plugging these functions into the equations of motion, one gets two first order differential equations for $F$ and $G$, and one second order differential equation for $J$. Then  there are four integration constants that can be fixed from: i)~asymptotically AdS$_5$ of the solution, i.e. $f(r)\to 1$ as $r\to\infty$; ii) regularity of $F(r)$ at the horizon; iii) vanishing of $A_\tau(r)$ at the horizon, or equivalently $J(r_h)=0$; and finally iV) the behavior of $A_{\tau}$ near the boundary must be $A_\tau(r) \sim A_{(0)\tau}\left(\frac{r}{\ell}\right)^{\Delta-3}$, cf. Eq.~(\ref{eq:Anb}). These conditions fix the constants and determine the temperature as a function of $r_h$ and $A_{(0)\tau}$, so that
\begin{equation}
T(r_h,A_{(0)\tau}) = T_0 \left( 1 + \frac{\delta T}{T_0} \right) \quad \textrm{where}  \qquad \frac{\delta T}{T_0} = F(r_h) + \frac{1}{2} G(r_h) \,.  \label{eq:Tr0}
\end{equation}
The explicit expressions for $f(r)$, $g(r)$ and $j(r)$ are too complicated to be shown here, but we will provide below the explicit results for the thermodynamical quantities which simplify a lot.

\subsection{Renormalization}
\label{subsec:renormalization}

In order to get the renormalized action, we need to add appropriate
counterterms to cancel the divergences of the action at the
boundary. In Fefferman-Graham coordinates the metric becomes as in
Eq.~(\ref{eq:metric}), while the gauge field is $A = A_{\tau}(\rho)
d\tau$ with $\rho = z^2$. From $g(r)$ one can~obtain the asymptotic
series giving $r$ in terms of $z$. The expansion near the boundary
turns out to be
\begin{equation}
r = \frac{\ell^2}{z} + a_1 z + a_3 z^3 + a_5 z^5 + b_5 z^5\log z + \cdots \,, \qquad z\to 0 \,,
\end{equation}
with the lowest order coefficient given by $a_1 = -q^2 m^2/(8 r_h^4)$. This series expansion yields explicit expressions for the metric components $g_{xx}(\rho)$ and $g_{\tau\tau}(\rho)$ near the boundary. After a straightforward computation, one can write the bulk contribution of the action as a total derivative, and then identify the counterterm needed to cancel the divergences. The result is
\begin{equation}
S_{\ct} = \frac{1}{2\kappa^2} \int d^4x \sqrt{h} \, \frac{6}{\ell} + \frac{\ell}{4 g_5^2} \int d^4x \sqrt{h} \, h^{ij} (\partial_i\Phi - m A_i) (\partial_j\Phi - m A_j) \,,
\end{equation}
where $h$ is the induced metric at the boundary. The renormalized action is then related to the thermodynamics of the system as
\begin{equation}
S_{\ren} = S_{\reg} + S_{\ct} = \frac{V_3}{T} \, w(r_h,q) \,, \label{eq:Sren}
\end{equation}
where $w$ is the free energy. In the last part of this section we will obtain explicit expressions for some of the thermodynamical variables.

\subsection{Thermodynamical variables}
\label{subsec:thermodynamical_variables}

These variables can be obtained from the thermodynamic potential Eq.~(\ref{eq:Sren}) by using the standard thermodynamical relations. At finite temperature and chemical potential, these read
\begin{equation}
p = -w \,, \qquad \varepsilon = - T^2 \frac{\partial}{\partial T}\left( \frac{w}{T} \right) \bigg|_{\mu} \,, \qquad s = -\frac{\partial w}{\partial T}  \bigg|_{\mu}  \,, \qquad  n = - \frac{\partial w }{\partial\mu} \bigg|_{T} \,, \label{eq:thermo_rel}
\end{equation}
where $n$ and $\mu$ are the charge density and chemical potential, respectively. These quantities fulfill the Euler identity~$\varepsilon + p = T s + \mu n$. Other quantity of interest is the speed of sound, which is defined as a variation at fixed entropy per particle, i.e.
\begin{equation}
c_s^2 = \left( \frac{\partial p}{\partial \varepsilon} \right)_{s/n} 
=   \left( \frac{\partial p}{\partial \varepsilon} \right)_{n} + \frac{n}{\varepsilon + p}  \left( \frac{\partial p}{\partial n} \right)_{\varepsilon} \,.
\end{equation}
Using these formulas and the free energy computed as explained in Sec.~\ref{subsec:renormalization}, one obtains that in the high temperature limit, $|A_{(0)\tau}| \ll T$, the trace anomaly and speed of sound write
\begin{equation}
\varepsilon - 3p  =  (\Delta-3) \frac{2\pi^2 \ell}{g_5^2} A_{(0)\tau}^2 T^2 + \cdots \,, \qquad c_s^2 = \frac{1}{3} - \frac{(\Delta-3) A_{(0)\tau}^2}{3\xi^2 T^2} + \cdots \,, \label{eq:thermo_highT}
\end{equation}
where $\xi = (\sqrt{3}\pi g_5 \ell)/(2\kappa)$, and the dots stand for
higher powers of $1/T$. Note that $\varepsilon - 3p \ge 0$ and $c_s^2
\le \frac{1}{3}$, so that the terms $\propto (\Delta-3)$ correspond to
non-conformal corrections induced by the deformation of the CFT,
cf. Sec.~\ref{subsec:Stueckelberg_model}. Finally, one can check the
validity of the Bekenstein-Hawking entropy formula in this
model. Using Eqs.~(\ref{eq:thermo_rel}) and (\ref{eq:thermo_highT})
one finds~$s = 2\pi r_h^3 / (\kappa^2 \ell^3)$. A trivial comparison
with the area of the event horizon, ${\cal A}$, computed with the
metric Eq.~(\ref{eq:dsBH}) leads to $s = 2\pi {\cal A} /\kappa^2$,
confirming the validity of the entropy relation. This means that the
usual laws of black hole thermodynamics are fulfilled also when
$\Delta > 3$, at least up to order $O(\Delta-3)$.

\section{Conclusions}
\label{sec:Conclusions}

In this work we have studied the thermodynamics of a holographic model
for QCD by applying techniques of singular perturbation theory. We
have seen that with these techniques the leading asymptotics of the
trace anomaly $\sim (\log T)^{-2}$ is resummed to give a dependence
controlled by the Lambert $\mathcal{W}$-function. This reproduces the
expected logarithmic suppression at high temperature, and it produces
an accurate description of the lattice data in the regime $T \geq 3
T_c$, even with a simple model having two parameters.

In the second part of this work we have studied the thermodynamical
properties of the St\"uckelberg model in 5D. We have obtained
analytical results by performing an expansion in the conformal
dimension of the current operator, and found that the expressions are
governed by pieces that break conformal invariance. The
Bekenstein-Hawking entropy formula turns out to be still valid, at
least at first order in the expansion.

\ack Part of this work is based on Ref.~\cite{Megias:2017czr},
co-authored with Manuel Valle. I would like to thank him for
collaboration and enlightening discussions. This research has been
supported by Spanish MINEICO and European FEDER funds (Grant
No. FIS2017-85053-C2-1-P), Plan Nacional de Altas Energ\'{\i}as
Spanish MINEICO (Grant No. FPA2015-64041-C2-1-P), Junta de
Andaluc\'{\i}a (Grant No. FQM-225), Basque Government (Grant
No. IT979-16), and Consejer\'{\i}a de Conocimiento, Investigaci\'on y
Universidad of the Junta de Andaluc\'{\i}a and European Regional
Development Fund (ERDF) (Grant No. SOMM17/6105/UGR), as well as by
Spanish MINEICO Ram\'on y Cajal Program (Grant No. RYC-2016-20678),
and by Universidad del Pa\'{\i}s Vasco UPV/EHU, Bilbao, Spain, through
a Visiting Professor appointment.

\section*{References}
%\begin{thebibliography}{9}
%\bibitem{iopartnum} IOP Publishing is to grateful Mark A Caprio, Center for Theoretical Physics, Yale University, for permission to include the {\tt iopart-num} \BibTeX package (version 2.0, December 21, 2006) with  this documentation. Updates and new releases of {\tt iopart-num} can be found on \verb"www.ctan.org" (CTAN). 
%\end{thebibliography}

\bibliography{refs}

\end{document}